\documentclass[twocolumn,showpacs,prb,superscriptaddress,floatfix]{revtex4}

\usepackage{color} 
\usepackage{tabularx} 
\usepackage{epsfig}
\usepackage{amsmath} 
\usepackage{amssymb} 
\usepackage{graphicx}
\usepackage{wasysym}
\usepackage{oldgerm}
\usepackage{psfrag}
\usepackage{algpseudocode}
\usepackage{color}    
\usepackage{subfigure}
\usepackage[usenames,svgnames]{xcolor}

\usepackage[utf8]{inputenc} 

\usepackage{tikz}
\usetikzlibrary{arrows,shapes,positioning}

\newcommand{\SU}{\mathrm{SU}(2)}
\newcommand{\LengthFunc}{{\rm len( )}}
\newcommand{\DistanceFunc}{{\rm d( )}}
\newcommand{\MatrixFunc}{{\rm mat( )}}
\newcommand{\LENGTH}[1]{{\rm len}({#1})}
\newcommand{\DISTANCE}[2]{{\rm d}({#1}, {#2})}
\newcommand{\MATRIX}[1]{{\rm mat}({#1})}
\newcommand{\ColorA}{Goldenrod}
\newcommand{\ColorB}{Red}
\newcommand{\ColorC}{DarkBlue}

\begin{document}

\title{Genetic braid optimization: A heuristic approach to compute
quasiparticle braids}

\author{Ross B.~McDonald}
\affiliation{Department of Physics and Astronomy, Texas A\&M University,
College Station, Texas 77843-4242, USA}

\author{Helmut G.~Katzgraber}
\affiliation{Department of Physics and Astronomy, Texas A\&M University,
College Station, Texas 77843-4242, USA}
\affiliation{Theoretische Physik, ETH Zurich, CH-8093 Zurich,
Switzerland}

\date{\today}

\begin{abstract}

In topologically protected quantum computation, quantum gates can be
carried out by adiabatically braiding two-dimensional quasiparticles,
reminiscent of entangled world lines. Bonesteel {\em et
al.}~[Phys.~Rev.~Lett.~{\bf 95}, 140503 (2005)], as well as Leijnse and
Flensberg [Phys.~Rev.~B {\bf 86}, 104511 (2012)], recently provided
schemes for computing quantum gates from quasiparticle braids.
Mathematically, the problem of executing a gate becomes that of finding
a product of the generators (matrices) in that set that approximates the
gate best, up to an error. To date, efficient methods to compute these
gates only strive to optimize for accuracy. We explore the possibility
of using a generic approach applicable to a variety of braiding problems
based on evolutionary (genetic) algorithms. The method efficiently finds
optimal braids while allowing the user to optimize for the relative
utilities of accuracy and/or length. Furthermore, when optimizing for error
only, the method can quickly produce efficient braids.

\end{abstract}


\maketitle

\section{Introduction}
\label{sec:intro}

Sensitivity to noise makes most of the current quantum computing schemes
prone to error and nonscalable, allowing only for small
proof-of-principle devices.  Topological quantum
computation\cite{freedman:02,kitaev:03} offers an elegant alternative to
overcome decoherence by using non-Abelian quasiparticles. Materials with
sufficiently complex topologically ordered phases can thus be used as
media for intrinsically fault-tolerant and scalable quantum information
processing. Different proposals and implementations exist to date,
ranging from fractional quantum Hall
systems\cite{moore:91,arovas:84,camino:05} with filling factors $\nu =
5/2$ and $\nu = 12/5$, conjectured to exhibit non-Abelian anyonic
excitations, \cite{bonderson:06,camino:06} to quantum dimer
models\cite{rokhsar:88,moessner:01,ioffe:02,misguich:02,fendley:05,ralko:05,moessner:03}
(implemented via Josephson junction
arrays),\cite{ioffe:02,albuquerque:08} spin and Hubbard
models,\cite{balents:02,freedman:05,sheng:05,isakov:06}
toric\cite{kitaev:03} and color\cite{bombin:06} codes, and an anisotropic spin
model,\cite{kitaev:06a} as well as implementations using cold
atomic\cite{buechler:05} or molecular\cite{micheli:06,pupillo:08} gases.
While the holy grail is the firm establishment of a phase with
non-Abelian anyons, virtually all aspects of topological quantum
computation are now under intense experimental and theoretical study.
Unfortunately, huge technological challenges remain for the development
of working devices; however, some proposals based on current
technologies exist.\cite{ioffe:02,albuquerque:08}

While bosons or fermions pick up phase factors of $\pm 1$ when braided,
for anyonic particles these simple phases are replaced by non-Abelian
matrices. The matrices act on a (degenerate) Hilbert space with a
dimensionality that grows exponentially in the number of quasiparticles
and whose states are intrinsically immune to decoherence because they
cannot be distinguished by local measurements.  A
topologically protected quantum gate then can be implemented by
adiabatically braiding
quasiparticles.\cite{freedman:02,kitaev:03,bonesteel:05,xu:11} At low
enough temperatures the system is, by design, protected from decoherence
because errors only can occur due to particle-hole excitations.

There are different quasiparticle systems that can be used to generate
quantum gates. However, in all cases the problem can be reduced to
finding a product of braid generators (matrices) that approximates a
given quantum gate with the smallest possible error and, if possible, as
short as possible. For example, in Bonesteel {\em et
al.}~\cite{bonesteel:05} braids of Fibonacci anyons\cite{freedman:02}
are computed. The fusion rules for these anyons make the Hilbert space
of the quasiparticles two-dimensional (see Sec.~\ref{sec:prob} for
details); i.e., a product of two-dimensional matrices has to be
computed. Bonesteel {\em et al.}~first performed a brute force
(exhaustive) search up to a braid length of 46 exchanges, obtaining
nontrivial gates with an error $\varepsilon \sim 10^{-3}$.
Unfortunately, the search space grows exponentially with the length of
the braid. Using the Solovay-Kitaev
algorithm,\cite{nielsen:00,dawson:06,hormozi:07} they then were able to
compute braids to any desired accuracy with a length that grows $\sim
|\log_{10}(\varepsilon)^4|$. For example, for an accuracy of
$\varepsilon \sim 10^{-3}$ the Solovay-Kitaev algorithm would require
braids of an approximate length of 81 exchanges. However, the
Solovay-Kitaev algorithm does not allow for the user to optimize for the
relative utilities of accuracy vs length. Depending on physical
implementations, a longer braid might be more problematic due to error
proliferation, and as such, having the option to either optimize for
accuracy and/or length might lead to braids better suited for a given
physical implementation.

In this paper we explore the possibility of using evolutionary (genetic)
algorithms\cite{hartmann:04} to efficiently find optimal braids while
allowing the user to optimize for the relative utilities of accuracy
and/or length.  We test the method with the braids computed by Bonesteel
{\em et al.},\cite{bonesteel:05} as well as a recent proposal by Leijnse
and Flensberg\cite{leijnse:12} that braids six Majorana fermions to
create two-qubit gates.  Furthermore, we show that when optimizing for
error only, the method can quickly produce efficient braids,
outperforming brute force searches.  We emphasize that the presented
method is generic and therefore can be applied to any problem that
requires the computation of the optimal product of (non-Abelian)
operators.  Thus it can be applied, for example, also to surface
codes.\cite{bocharov:12}

This paper is structured as follows. In Sec.~\ref{sec:prob} we show how
the complex quantum computing problem can be reduced to a simple
mathematical problem of finding matrix products, followed by a brief
review of previous methods in Sec.~\ref{sec:previous}.  The evolutionary
algorithm is introduced in Sec.~\ref{sec:algo}, followed by results in
Sec.~\ref{sec:results} and concluding remarks.

\section{Simplified Problem Representation}
\label{sec:prob}

In this section we illustrate the method on two different
quantum computing proposals.

A braid operation can be represented by a matrix that acts on the qubit
space. These matrices will be referred to as {\em generators}, and the
quantum gate that a braid represents is the product of the generators
that represent the individual braid operations. The problem of finding
braiding operations that approximate gates is then reduced to finding a
product chain of the reduced generators and their inverses that
approximates the matrix representing the quantum gate.

Fibonacci anyon braids\cite{bonesteel:05} only encompasses one-qubit gates.
In such systems, the braid transition operators result in a phase change
for the noncomputational state, and therefore it can be ignored. Overall,
phases in the problem can also be ignored. Therefore the transition
matrices can be projected onto $\SU$ by a multiplication with $e^{\imath
\pi / 10}$, yielding for the generators and their graphical
representations
\begin{center}
\begin{tabular}{ c c c c c }
$\sigma_1$
& $=$
& $\begin{pmatrix}
e^{-\imath 7\pi/10} & 0 \\
0 & -e^{-\imath 3\pi/10} 
\end{pmatrix}$
& $=$
& \raisebox{-0.45\height}{\begin{tikzpicture}
\begin{scope}[line width=2pt]
\draw[\ColorA] (0,0.0) -- +(0.35,0);
\draw[\ColorB] (0,-0.8) .. controls +(0:0.175) and +(180:0.175) .. (0.35,-0.4);
\draw[\ColorC] (0,-0.4) .. controls +(0:0.175) and +(180:0.175) .. (0.35,-0.8);
\end{scope}
\end{tikzpicture}
} \\
$ \sigma_2$
& $=$
& $\begin{pmatrix}
-\tau e^{-\imath \pi/10} & -\imath \sqrt{\tau} \\
-\imath \sqrt{\tau} & -\tau e^{\imath \pi/10}
\end{pmatrix}$
& $=$
& \raisebox{-0.45\height}{\begin{tikzpicture}
\begin{scope}[line width=2pt]
\draw[\ColorA] (0,-0.8) -- +(0.35,0);
\draw[\ColorB] (0,-0.4) .. controls +(0:0.175) and +(180:0.175) .. (0.35,0.0);
\draw[\ColorC] (0,0.0) .. controls +(0:0.175) and +(180:0.175) .. (0.35,-0.4);
\end{scope}
\end{tikzpicture}
} \\
\end{tabular}
\end{center}
where $\tau = (\sqrt{5}-1)/2$, and the graphical representations are
those used, for example, in Fig.~\ref{fig:cool-braid}.

In the Leijnse and Flensberg scheme based on Majorana fermions the braid
operators act on a two-qubit system; i.e., the gates will be two-qubit
gates. The generators for this scheme are higher-dimensional, i.e.,
\begin{widetext}
\begin{eqnarray}
B_1 = 
\begin{pmatrix}
\imath & 0 & 0 & 0 \\
0 & \imath & 0 & 0 \\
0 & 0 & 1 & 0 \\
0 & 0 & 0 & 1 
\end{pmatrix}
, \;
B_2 = \frac{1}{\sqrt{2}}
\begin{pmatrix}
1 & 0 & \imath & 0 \\
0 & 1 & 0 & \imath \\
\imath & 0 & 1 & 0 \\
0 & \imath & 0 & 1
\end{pmatrix}
, \;
B_3 =
\begin{pmatrix}
\imath & 0 & 0 & 0 \\
0 & 1 & 0 & 0 \\
0 & 0 & 1 & 0 \\
0 & 0 & 0 & \imath
\end{pmatrix}
, \;
B_4 = \frac{1}{\sqrt{2}} 
\begin{pmatrix}
1 & \imath & 0 & 0 \\
\imath & 1 & 0 & 0 \\
0 & 0 & 1 & -\imath \\
0 & 0 & -\imath & 1 
\end{pmatrix}
, \;
B_5 = 
\begin{pmatrix}
\imath & 0 & 0 & 0 \\
0 & 1 & 0 & 0 \\
0 & 0 & \imath & 0 \\
0 & 0 & 0 & 1
\end{pmatrix} .
\end{eqnarray}
\end{widetext}
The goal is now to find a product of generator matrices that produces a
braid that represents a gate operation under the constraints that either
length is minimized, error is minimized, or both length and error are
minimized.

\section{Traditional Approaches}
\label{sec:previous}

\begin{figure}
\includegraphics[width=1.00\columnwidth]{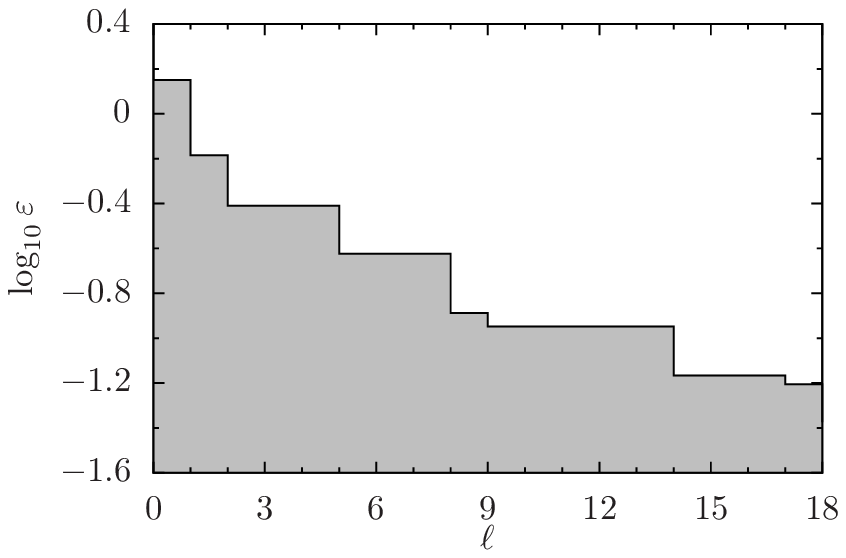}
\caption{
Minimum possible error $\varepsilon$ for a braid of maximum length
$\ell$ of Fibonacci anyons. This brute-force result for braids up to
$\ell = 18$ is a lower bound for the error for all other search methods.
The line corresponds to the optimum; i.e., any heuristic method can only
produce a solution that lies on or above the line.
\label{fig:brute-force}}
\end{figure}

The na\"{i}ve approach to solve the braiding problem is a brute-force
search.  A target error is set, and the set of all braids is searched
from shortest to longest until a braid whose error is smaller than or
equal to the target error is found.  However, this approach is
nonscalable, as illustrated in Fig.~\ref{fig:brute-force} for Fibonacci
anyons. In this case we have four possible matrices (two generators and
their inverses) for each position in the braid (two or three choices if
cancellations between inverses are not ignored). This means that the
number of different braids of length $\ell$ is $4^\ell$ (or in the range
$2^\ell$ -- $3^\ell$ including cancellations).  This is even worse in
the Leijnse and Flensberg scheme for Majorana fermions where one has 10
different matrices; i.e., an exhaustive search for a braid of length
$\ell$ might have a worst-case run time of order $O(10^\ell)$.  Because
the number of possible braids grows exponentially with length, a
brute-force search would be too slow for most practical applications.
Note, however, that a bidirectional
search\cite{xu:08,hormozi:09,zikos:09} greatly improves the performance.

The Solovay-Kitaev algorithm provides a boost in the efficiency of
finding more accurate braids, but at the cost of accepting braids that
are longer than necessary. Depending on the implementation, this might
be problematic: While a given error might be desirable, a given hardware
implementation might degrade considerably with the length of the braid.
In such cases short braids might be desirable.  Given a target error of
$\varepsilon$, the Solovay-Kitaev algorithm produces braids of length
$O[\log_{10}^{3.97}(1/\varepsilon)]$ that are guaranteed to have an
error less than $\varepsilon$ in a run time of
$O[\log_{10}^{2.71}(1/\varepsilon)]$.\cite{dawson:06,hormozi:07}

Another option explored by Burrello {\em et al}.\cite{burrello:10} is
braid hashing, in which approximations of the identity braid are used to
refine crude approximations of the target braid into more accurate
solutions. This method is fast and can produce very accurate braids, but
it does not address the problem of increasing braid length with
accuracy. Furthermore, the braid hashing algorithm works only for
Fibonacci anyons, and it is unclear how the method can be generalized to
other systems, such as Majorana fermions.

In particular, none of these methods seek to optimize braid length in
addition to accuracy except for brute force; however, brute force is
slow. Below we present an efficient algorithm that can be tuned to
optimize for both length and/or accuracy. The method has the potential
to create shorter braids than the Solovay-Kitaev algorithm.  However, we
note that the Solovay-Kitaev algorithm is faster and can overcome some
convergence problems the genetic approach faces (see below). In
comparison to brute-force methods, the genetic approach yields good
results considerably faster. We also emphasize that this method is
applicable to any system where quantum gates are built from a finite set
of fault-tolerant gates. Although there have been some attempts to solve
this problem generically, the approach of, e.g.,
Ref.~\onlinecite{bocharov:12} only applies to single-qubit systems. The
genetic method outlined below can be potentially applied to arbitrary
systems.

\section{Evolutionary Algorithm}
\label{sec:algo}

Mathematically, the problem at hand is similar to solving a Rubik's
cube: The goal is to find the shortest set of matrix operations (cube
rotations) to obtain the minimum of a cost function (uniformly colored
faces of the cube), which here represents the shortest braid or the
smallest error.

The proposed algorithm resembles a steady-state genetic algorithm: A
population of random-solution braids is generated in the initialization
step of the algorithm.  The population then evolves in an iterative
process where different generations are developed according to
predefined mathematical operations on the population.  Each update on a
generation is broken into two steps, \textit{culling} and
\textit{breeding}, which are described in detail below.  After a predefined
number of generations have been executed, the algorithm terminates, and
the best braid encountered by the algorithm is presented as the
solution. A ``best braid'' is defined as the braid with the highest
fitness, i.e., the braid that minimizes the problem-dependent cost
function (described below) for the problem.

To simplify matters, we introduce the following notation: Let
$B^{[a,b]}$ denote the sub braid of braid $B$ from the $a$th element to
$b$th element (inclusive), and let $B^{[a,]}$ denote the sub braid of
$B$ from the $a$th element to the end of $B$. Furthermore, let the
concatenation of braid variables indicate a concatenation of the actual
braids. For example, $B=B_1 B_2$ means that $B$ is a concatenation of
braids $B_1$ and $B_2$.  $B=B_1^{[3,]} B_2^{[2,5]}$ would indicate that
$B$ is a concatenation of the third to end sub braids of $B_1$ and the
second to fifth sub braids of $B_2$.  Let $\LengthFunc$ be a function
that has a braid as its argument and returns that braid's length, let
$\MatrixFunc$ be a function that has a braid as its argument and returns
the product of the braid's elements in matrix form, and let
$\DistanceFunc$ be a function that evaluates the distance between two
braids, specifically,
\begin{equation}
\DISTANCE{B_1}{B_2} = \left| \MATRIX{B_1} - \MATRIX{B_2} \right| .
\label{eq:metric}
\end{equation}
Here and for the rest of the paper, the matrix norm used is
\begin{equation}
|X| = \sqrt{\sum_{ij} X_{ij}^2}.
\end{equation}  
Equation \eqref{eq:metric} defines the metric used to determine the
``distance'' between two braids.

A high-level view of the simulation is provided by the following
pseudocode in which $m$ is the population size, $generations$ is the
number of generations to evolve the population, and $best$ is the
current best braid:
\begin{algorithmic}
	\State $population \gets$ new population of size $m$
	\For{$i = 1 \to generations$}
		\State Sort $population$ ascending by fitness
		\If{$fitness(best) < fitness(population[m])$}
			\State $best \gets population[m]$
		\EndIf
		\State Perform culling (least fit 10\% removed)
		\State Repopulate missing 10\% with breeding
		\State $i \gets i+1$
	\EndFor
\end{algorithmic}
It should be noted that in this pseudocode and all following
pseudocodes, collections use one-based indices (i.e., $population[1]$ is
the first element of population).  At the end of a generation the
population is the same size as in the beginning; however, we expect that
the offsprings are fitter than the initial randomly chosen parents.

\subsection{Culling}
\label{subsec:culling}

The population is sorted according to a fitness function. The fitness
can, in general, be any real-valued function of braid length $\ell$ and
error $\varepsilon$.  Here we use
\begin{equation}
f(\ell, \varepsilon) = 
\frac{1-\lambda}{1+\varepsilon} + \frac{\lambda}{\ell} .
\end{equation}
The braid error is calculated with the following metric:
\begin{equation}
\varepsilon = \left|B - X \right| , 
\end{equation}
where $B$ is a braid matrix and $X$ represents the target matrix (gate
to be emulated). The parameter $\lambda$ allows one to tune between a
short braid or a more accurate braid; i.e., for $\lambda \to 1$ the
system is tuned for length, whereas for $\lambda \to 0$ the system is
tuned for error reduction.

After sorting the population by fitness, the bottom 10\% of the genes
(braids) are removed.

\subsection{Breeding}
\label{subsec:breeding}

After a culling operation only 90\% of the genes are left in the gene
pool.  The remaining 10\% are filled by combining the remaining braids into
new braids, i.e., breeding.  From the 90\% of the braids that survived
the culling operation, which represents the top 90\% of the genes in
the population, two braids are selected as parents of a new braid for
the gene pool. (Note that values $\sim$90\% are typically used in the
literature). Let these parents be denoted as $P_1$ and $P_2$, and let
the offspring be denoted as $C_1$ and $C_2$.  The way two parent braids
are combined into an offspring plays a {\em crucial} role in the
efficiency of the algorithm.

Our initial rather na\"{i}ve approach was to select two random
integers $n_1$ and $n_2$ such that $n_1 \in (1,\LENGTH{P_1}]$ and $n_2
\in (1,\LENGTH{P_2}]$. The boundaries of these ranges are chosen to
prevent duplication of the parents. The offspring are then formed as
$C_1 = P_1^{[1,n_1 -1]} P_2^{[n_2,]}$ and $C_2 = P_2^{[1,n_2-1]}
P_1^{[n_1,]}$.  Due to the noncommutativity of matrix multiplication,
this method is no better than {\em randomly generating} two offspring.
To remedy this problem, we used a different recombination method,
referred to as {\em contextual recombination}. In contextual
recombination the partition points $n_1$ and $n_2$ are chosen by
minimizing the distance between the first halves of the two parent
braids. However, to avoid cloning, we require that the first halves not
be identical. The actual recombination method after $n_1$ and $n_2$ are
chosen is the same as above. To choose $n_1$ and $n_2$, one must first
determine $m$, where $m$ is the largest integer such that $P_1^{[1,m]}$
and $P_2^{[1,m]}$ both exist and are identical. Once $m$ is determined,
$n_1$ and $n_2$ are chosen such that $n_1 \in (m, \LENGTH{P1}]$, $n_2
\in (m, \LENGTH{P_2}]$, and $\DISTANCE{P_1^{[1, n_1 - 1]}}{P_2^{[1, n_2
-1]}}$ is minimized.  These values can be determined using the following
pseudocode:
\begin{algorithmic}
\State $minDistance \gets \infty$
\State $m \gets 0$
\While{$m < min(\LENGTH{P_1}, \LENGTH{P_2})$ and $P_1^{[m,m]} = P_2^{[m,m]}$}
	\State $m \gets m+1$
\EndWhile
\For{$i=m \to \LENGTH{P_1}$}
       \For{$j=m \to \LENGTH{P_2}$}
               \State $dist \gets \DISTANCE{P_1^{[1, i-1]}}{P_2^{[1, j-1]}}$
               \If{$dist < minDistance$}
                       \State $minDistance \gets dist$
                       \State $n_1 \gets i$
                       \State $n_2 \gets j$
               \EndIf
       \EndFor
\EndFor
\end{algorithmic}

\section{Results}
\label{sec:results}

We have run the algorithm for both Fibonacci anyons and Majorana
particles.  The population sizes are 80 individuals for both cases
following the recommendations by Belmont-Moreno.\cite{belmont-moreno:01}
Increasing the population size showed no significant improvement on the
results but increased the run time.

We start with Fibonacci anyons as studied in
Ref.~\onlinecite{bonesteel:05} with $\lambda=0$.  In this case we only
optimize for accuracy and not for length. The goal is to emulate the
$X$-rotation gate
\begin{equation}
X = 
\begin{pmatrix}
0 & \imath \\
\imath & 0
\end{pmatrix}.
\end{equation}
In the two-qubit case we emulate the controlled NOT (or CNOT) gate, namely
\begin{equation}
X =
\begin{pmatrix}
1 & 0 & 0 & 0 \\
0 & 1 & 0 & 0 \\
0 & 0 & 0 & 1 \\
0 & 0 & 1 & 0
\end{pmatrix}.
\end{equation}

\begin{figure}
\includegraphics[width=1.00\columnwidth]{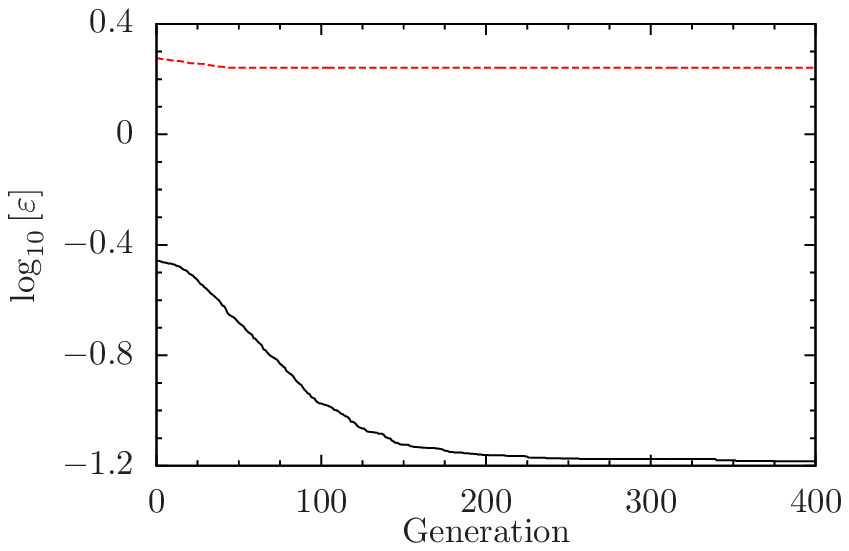}
\caption{(Color online)
Average error in the population for each generation averaged over 100
sample runs of the algorithm for $\lambda=0$. The solid line represents
the average error for Fibonacci anyons when emulating an $X$-rotation gate. 
The dashed line is for Majorana particles emulating a CNOT gate.
\label{fig:lambda-zero}}
\end{figure}

Figure \ref{fig:lambda-zero} shows the error $\varepsilon$ as a function
of the run time of the algorithm. The method is capable of improving
solutions very quickly (after approximately 150 generations) but is
unable to improve after a certain point. An interesting example braid of
the algorithm that shows its potential can be seen in
Fig.~\ref{fig:cool-braid}. This is very promising, and we believe the
minimum error problem can be solved by introducing a clever mutation
method.  However, our attempts to implement a basic mutation (i.e.,
changing a generator into a random new one and inserting approximations
to the identity braid into the braid) rendered the algorithm as
inefficient as a random search.

Unfortunately, for two-qubit gates (Majorana fermions) the algorithm is
not efficient. Figure \ref{fig:lambda-zero} shows the error
$\varepsilon$ as a function of the run time (dashed line). The data
converge quickly to a plateau and cease to improve; i.e., the accuracy of
the braid cannot be improved.

\begin{figure}
\begin{tikzpicture}
\begin{scope}[line width=2pt]
\draw[\ColorC] (0,-0.8) -- +(0.35,0);
\draw[\ColorA] (0,0.0) .. controls +(0:0.175) and +(180:0.175) .. (0.35,-0.4);
\draw[\ColorB] (0,-0.4) .. controls +(0:0.175) and +(180:0.175) .. (0.35,0.0);
\draw[\ColorC] (0.35,-0.8) -- +(0.35,0);
\draw[\ColorB] (0.35,0.0) .. controls +(0:0.175) and +(180:0.175) .. (0.7,-0.4);
\draw[\ColorA] (0.35,-0.4) .. controls +(0:0.175) and +(180:0.175) .. (0.7,0.0);
\draw[\ColorA] (0.7,0.0) -- +(0.35,0);
\draw[\ColorC] (0.7,-0.8) .. controls +(0:0.175) and +(180:0.175) .. (1.05,-0.4);
\draw[\ColorB] (0.7,-0.4) .. controls +(0:0.175) and +(180:0.175) .. (1.05,-0.8);
\draw[\ColorA] (1.05,0.0) -- +(0.35,0);
\draw[\ColorB] (1.05,-0.8) .. controls +(0:0.175) and +(180:0.175) .. (1.4,-0.4);
\draw[\ColorC] (1.05,-0.4) .. controls +(0:0.175) and +(180:0.175) .. (1.4,-0.8);
\draw[\ColorA] (1.4,0.0) -- +(0.35,0);
\draw[\ColorC] (1.4,-0.8) .. controls +(0:0.175) and +(180:0.175) .. (1.75,-0.4);
\draw[\ColorB] (1.4,-0.4) .. controls +(0:0.175) and +(180:0.175) .. (1.75,-0.8);
\draw[\ColorA] (1.75,0.0) -- +(0.35,0);
\draw[\ColorB] (1.75,-0.8) .. controls +(0:0.175) and +(180:0.175) .. (2.1,-0.4);
\draw[\ColorC] (1.75,-0.4) .. controls +(0:0.175) and +(180:0.175) .. (2.1,-0.8);
\draw[\ColorC] (2.1,-0.8) -- +(0.35,0);
\draw[\ColorA] (2.1,0.0) .. controls +(0:0.175) and +(180:0.175) .. (2.45,-0.4);
\draw[\ColorB] (2.1,-0.4) .. controls +(0:0.175) and +(180:0.175) .. (2.45,0.0);
\draw[\ColorB] (2.45,0.0) -- +(0.35,0);
\draw[\ColorC] (2.45,-0.8) .. controls +(0:0.175) and +(180:0.175) .. (2.8,-0.4);
\draw[\ColorA] (2.45,-0.4) .. controls +(0:0.175) and +(180:0.175) .. (2.8,-0.8);
\draw[\ColorA] (2.8,-0.8) -- +(0.35,0);
\draw[\ColorB] (2.8,0.0) .. controls +(0:0.175) and +(180:0.175) .. (3.15,-0.4);
\draw[\ColorC] (2.8,-0.4) .. controls +(0:0.175) and +(180:0.175) .. (3.15,0.0);
\draw[\ColorC] (3.15,0.0) -- +(0.35,0);
\draw[\ColorA] (3.15,-0.8) .. controls +(0:0.175) and +(180:0.175) .. (3.5,-0.4);
\draw[\ColorB] (3.15,-0.4) .. controls +(0:0.175) and +(180:0.175) .. (3.5,-0.8);
\draw[\ColorB] (3.5,-0.8) -- +(0.35,0);
\draw[\ColorA] (3.5,-0.4) .. controls +(0:0.175) and +(180:0.175) .. (3.85,0.0);
\draw[\ColorC] (3.5,0.0) .. controls +(0:0.175) and +(180:0.175) .. (3.85,-0.4);
\draw[\ColorA] (3.85,0.0) -- +(0.35,0);
\draw[\ColorC] (3.85,-0.4) .. controls +(0:0.175) and +(180:0.175) .. (4.2,-0.8);
\draw[\ColorB] (3.85,-0.8) .. controls +(0:0.175) and +(180:0.175) .. (4.2,-0.4);
\draw[\ColorA] (4.2,0.0) -- +(0.35,0);
\draw[\ColorB] (4.2,-0.4) .. controls +(0:0.175) and +(180:0.175) .. (4.55,-0.8);
\draw[\ColorC] (4.2,-0.8) .. controls +(0:0.175) and +(180:0.175) .. (4.55,-0.4);
\draw[\ColorB] (4.55,-0.8) -- +(0.35,0);
\draw[\ColorC] (4.55,-0.4) .. controls +(0:0.175) and +(180:0.175) .. (4.9,0.0);
\draw[\ColorA] (4.55,0.0) .. controls +(0:0.175) and +(180:0.175) .. (4.9,-0.4);
\draw[\ColorC] (4.9,0.0) -- +(0.35,0);
\draw[\ColorA] (4.9,-0.4) .. controls +(0:0.175) and +(180:0.175) .. (5.25,-0.8);
\draw[\ColorB] (4.9,-0.8) .. controls +(0:0.175) and +(180:0.175) .. (5.25,-0.4);
\draw[\ColorA] (5.25,-0.8) -- +(0.35,0);
\draw[\ColorC] (5.25,0.0) .. controls +(0:0.175) and +(180:0.175) .. (5.6,-0.4);
\draw[\ColorB] (5.25,-0.4) .. controls +(0:0.175) and +(180:0.175) .. (5.6,0.0);
\draw[\ColorA] (5.6,-0.8) -- +(0.35,0);
\draw[\ColorB] (5.6,0.0) .. controls +(0:0.175) and +(180:0.175) .. (5.95,-0.4);
\draw[\ColorC] (5.6,-0.4) .. controls +(0:0.175) and +(180:0.175) .. (5.95,0.0);
\draw[\ColorA] (5.95,-0.8) -- +(0.35,0);
\draw[\ColorC] (5.95,0.0) .. controls +(0:0.175) and +(180:0.175) .. (6.3,-0.4);
\draw[\ColorB] (5.95,-0.4) .. controls +(0:0.175) and +(180:0.175) .. (6.3,0.0);
\draw[\ColorA] (6.3,-0.8) -- +(0.35,0);
\draw[\ColorB] (6.3,0.0) .. controls +(0:0.175) and +(180:0.175) .. (6.65,-0.4);
\draw[\ColorC] (6.3,-0.4) .. controls +(0:0.175) and +(180:0.175) .. (6.65,0.0);
\draw[\ColorA] (6.65,-0.8) -- +(0.35,0);
\draw[\ColorC] (6.65,0.0) .. controls +(0:0.175) and +(180:0.175) .. (7.0,-0.4);
\draw[\ColorB] (6.65,-0.4) .. controls +(0:0.175) and +(180:0.175) .. (7.0,0.0);
\draw[\ColorB] (7.0,0.0) -- +(0.35,0);
\draw[\ColorA] (7.0,-0.8) .. controls +(0:0.175) and +(180:0.175) .. (7.35,-0.4);
\draw[\ColorC] (7.0,-0.4) .. controls +(0:0.175) and +(180:0.175) .. (7.35,-0.8);
\draw[\ColorC] (7.35,-0.8) -- +(0.35,0);
\draw[\ColorB] (7.35,0.0) .. controls +(0:0.175) and +(180:0.175) .. (7.7,-0.4);
\draw[\ColorA] (7.35,-0.4) .. controls +(0:0.175) and +(180:0.175) .. (7.7,0.0);
\end{scope}
\end{tikzpicture}
$$ = \sigma_2^{-2} \sigma_1^4 \sigma_2^{-1} \sigma_1 \sigma_2^{-1} \sigma_1
\sigma_2 \sigma_1^{-2} \sigma_2 \sigma_1^{-1} \sigma_2^{-5} \sigma_1
\sigma_2^{-1} \approx 
\begin{pmatrix}
0 & \imath \\ \imath & 0
\end{pmatrix}
$$
\caption{(Color online)
Example result for a braid of Fibonacci anyons emulating the
$X$-rotation gate.  The error of this approximation is $3.1 \times
10^{-3}$.
\label{fig:cool-braid}}
\end{figure}

\begin{figure}
\includegraphics[width=1.00\columnwidth]{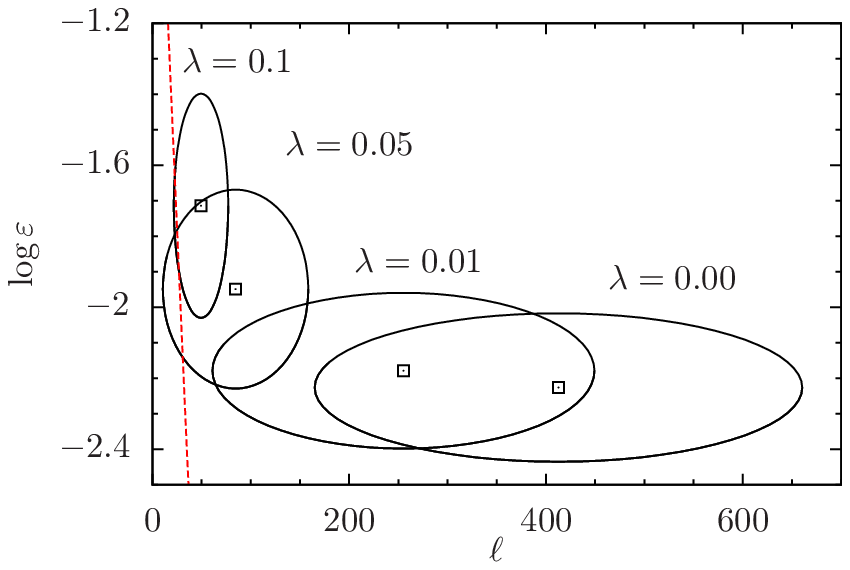}
\caption{(Color online)
Distribution of the output of the algorithm for different values of
$\lambda$.  Each ellipse represents the output distribution for a single
value of $\lambda$ centered on the average output, with the ellipse
bounds being a single standard deviation from the mean.
Averages are over 100 runs.  The dashed red line is an extrapolation of
the graph in Fig.~\ref{fig:brute-force}.
\label{fig:lambdas}}
\end{figure}

Figure \ref{fig:lambdas} shows that by tuning $\lambda$ in the fitness
function we are able to tune fitness against accuracy effectively for
the case of Fibonacci anyons emulating the $X$-rotation gate.  The squares
represent averages over 100 runs and the ellipses correspond to standard
deviations.  The variance for small $\lambda$ is large. However, for
larger values of $\lambda$ the length of the braid can be effectively
constrained. Although the spread in the accuracy is large, repeating the
simulation multiple times allows one to determine an optimal braid with
a small error and small length quickly (less than 1h for 100 runs on an
average CPU).  We expect that by introducing clever mutations the spread
in the data can be reduced.

Figure \ref{fig:lambdas} also shows that for high values of $\lambda$,
the algorithm produces results very near the best-case boundary but is
constrained to the high-error region.  As $\lambda$ decreases, the
solutions move away from the best-case boundary, producing
longer-than-needed braids.

\section{Conclusions}
\label{sec:conclusions}

We have introduced a {\em generic} algorithm based on evolutionary
methods to approximate gates using quasiparticle braids.  While
single-qubit braids of Fibonacci anyons can be computed efficiently, the
method fails to produce optimal braids for two-qubit gates. The latter
presents an unresolved challenge that we will attempt to tackle in the
near future. Our results suggest that mutations might be key in the
improvement of the method.

We emphasize that the developed method is generic and therefore can be
applied to problems ranging from general quantum compiling, to orienting
devices using coarse stepper motors in industrial applications, as well
as generic optimization of problems with competing goals.  It would be
interesting to compare our results to bidirectional
search,\cite{xu:08,hormozi:09,zikos:09} which we plan to do in the
future.

\begin{acknowledgments} 

We would like to thank N.~Bonesteel for useful discussions.
H.G.K.~acknowledges support from the Swiss National Science Foundation
(Grant No.~PP002-114713) and the National Science Foundation (Grant
No.~DMR-1151387).

\end{acknowledgments}

\bibliography{refs}
 
\end{document}